\documentclass[prl,aps,twocolumn,showpacs]{revtex4-1} 
\usepackage{graphicx}
\begin{document}

\def\ii{\'\i}
\def \st {\widetilde \sigma}
\def \sd {\sigma ^{\dag}}
\def \pt {\widetilde\pi}
\def \pd {\pi ^{\dag}}
\def \fn {$\Phi_{NLM}$}
\def \fav {$\Phi_{A^{'}}$}
\def \fx {$\Phi _{x}$}
\def \fa {$\Phi _{A}$}
\def \np{n_{\pi}}
\def \f7 {$f_{7/2}$}
\def \si {$\sigma$}
\def \oa {$^{16}O + \alpha$ \ }
\def \ca {$^{12}C + \alpha$ \ }
\def \cam {$^{12}C + \alpha$}
\def \co {$^{12}C + \ ^{16}O$ \ }
\def \com {$^{12}C + \ ^{16}O$}
\def \ma {$^{24}Mg + \alpha$}
\def \mam {$^{24}Mg + \alpha$ \ }
\def \cc {$^{12}C + \ ^{12}C$ \ }
\def \ust {$U^{ST}(4)$ \ }
\def \ustc {$U^{ST}_C(4)$ \ }

\pagebreak
\title{On the intersection of the shell, collective and cluster models of atomic nuclei I:
Multi-shell excitations}
\author{J.~Cseh}
\affiliation{Institute for Nuclear Research, Hungarian Academy of Sciences, Debrecen, Pf. 51, Hungary-4001}
\date{\today}

\begin{abstract}
The relation of the shell, collective and cluster models of the atomic nuclei
is discussed from the viewpoint of symmetries. 
In the fifties the U(3) symmetry  was found as their common part  for a single shell problem.
For multi major-shell excitations, considered here, the U(3)$\otimes$U(3) dynamical symmetry
turns out to be their intersection.

\end{abstract}

\pacs{21.10.Re, 21.60.Cs, 21.60.Fw, 21.60.Gx, 27.30.+t}    
\maketitle

\section{Introduction}

The fundamental models of nuclear structure are based on different physical pictures.
The shell model indicates that the atomic nucleus is something like a small atom,
the cluster model suggests that it is similar to a molecule, while 
the collective model says that it is a microscopic liquid drop.
Therefore, in order to  understand  the nuclear structure we need to
study  (among others) the interrelation of these models, find their common intersection, etc.

The basic connections were found in the fifties. Elliott 
\cite{elli58}
 showed
how the quadrupole deformation and collective rotation can be derived
from the spherical shell model: the states belonging to a collective band are determined
by their specific SU(3) symmetry. 
Wildermuth and Kanellopoulos 
\cite{wika58}
(following  Perring and Skyrme 
\cite{foll})
established the relation between the shell and cluster models. They proved that
the Hamiltonians of the two models can be rewritten into each other exactly in the
harmonic oscillator approximation. This relation results in a close connection between the
corresponding eigenvectors, too: the wavefunction of one model is a linear combination of
those of the other, which belong to the same energy. Later on this relation was interpreted
by Bayman and Bohr 
\cite{babo58}
in terms of the SU(3) symmetry. As a consequence, the cluster 
states are also selected from the shell model space by their specific
SU(3)  symmetries.
(In fact only one kind of cluster states have this feature, and there are other kinds, too,
as discussed later.)

We will refer to this interrelation among the three basic structure  models as the SU(3)
connection. It was established in 1958  for a single major shell problem, 
for simple symmetries and small deformations of the ground-state region.
In this paper we
consider its extension to multi-major-shells,
and in a following one the more general symmetries and the 
case of large deformations are considered
\cite{csdII}.
The concept is illustrated by the example of the $^{20}$Ne nucleus.

The connection of the shell model and the cluster model is especially interesting
due to the fact that both models have a complete set of basis states, i.e.
any nuclear states can be expanded in both bases. Depending on the  simple
or complicated nature of these expansions we can distinguish four different
cases. i) If it is simple in the shell basis, but complicated in the cluster one,
then we can speak about a simple shell state, which is a poor cluster state.
ii) If it is simple in the cluster basis, but complicated in the shell basis, one
has a good cluster state, which is a poor shell state. (Later on we will refer
to these states as rigid molecule-like cluster states.) 
iii) The expansion  may be simple in
both bases. We call this situation as a shell-like cluster state. iv) If it is 
complicated in  both bases, than the state is not a simple one.

Of course it is not the name that matters. The important point is that
there are three different kinds of simple states when we investigate them
from the viewpoint of the cluster-shell competition. Sometimes only the
kind ii), i.e. rigid molecule-like states are considered as cluster states.
The reason for our vocabulary, i.e. speaking about two kinds of
cluster states: shell-like and molecule-like is inspired by their experimentally
observable characteristics. Both of them prefer a certain reaction channel, have
a large cluster spectroscopic factor, etc. 
They represent two different quantum phases of the nucleus,
as discussed in
\cite{csdII}.
By choosing these names we
are in line with the general definition of a simple nuclear state in terms of
experimental observation: a state is called simple if its wavefunction has a
large overlap, or form a large matrix element with the wavefunction of a reaction channel,
in which it can be observed
\cite{maho}.

The structure of this paper is as follows.
First we review some aspects of the shell, collective and cluster models, which are
relevant for their symmetry-based relations, then we discuss their intersection.

\section{Shell models}
\subsection{Single major shell}

Elliott's U(3) model bridges the spherical shell model and the collective model,
and describes rotational states in the $p$ and $sd$ shell nuclei
\cite{arim99}.
Here the U(3) symmetry is that of the space part, while the
spin-isospin section  is characterized by Wigner's U$^{ST}$(4) 
\cite{wig37}. Thus the group structure of the model is
U$^{ST}$(4) $\otimes$ U(3).
In constructing the model space the spin-isospin degrees of freedom are
essential, of course. 
The physical operators, however,
often contain simply spin-isospin zero terms,
in which case  only the generators of the U(3) group contribute.

For a single-particle problem
the  U(3) group is generated by the oscillator quantum creation 
($\pd_{\mu} ,  \mu = -1,0,+1$ ) and annihilation
($ \pi_{\mu} $)
operators, which carry $l=1$ angular momentum,
i.e. they are vector bosons. Their 9  number-conserving 
bilinear products can be recast into a scalar 
number operator, the three components of the angular momentum,
and the five components of the (algebraic, i.e. no major shell changing) 
quadrupole operator, as follows:
\begin{eqnarray}
n  =
\sqrt {3}
[\pi ^{\dag} \times \widetilde \pi ]^{(0)}_0 , \
L_{m}  =
\sqrt {2}
[\pi ^{\dag} \times \widetilde \pi ]^{(1)}_m ,
\nonumber \\
Q_{m}  =
{{\sqrt {3}}/2}
[\pi ^{\dag} \times \widetilde \pi ]^{(2)}_m \ ,
 \label{sumgen}
\end{eqnarray}
where
$
\widetilde {\pi}_{ m} = (-1)^{1 - m} {\pi}_{ -m} \ .
$
Removing the number operator $n$ the remaining 8 operators
generate the SU(3) subgroup. 
For the $A$-nucleon system the operators are obtained by summing on
the particle number:
$
O = {\Sigma}_{i=1}^{A}O(i),
$
but  the Equations 
(\ref{sumgen}) remain valid for the many-body operators, too.

The space part of the basis states is defined by the
representation labels of the group chain:
\begin{eqnarray}
 U(3) \supset SU(3) \supset SO(3) \supset SO(2)
 \nonumber \\
\vert [n_1,n_2,n_3]  ,  (\lambda , \mu) ,\  K\ ,  \ \ L \ \ \ \ ,\ \ \  M  \ \rangle .
 \label{eq:ellgrch}
\end{eqnarray}
Here $ n = n_1 +n_2 +n_3 , \ \lambda = n_1 - n_2,\  \mu = n_2 - n_3$, 
$K$ distinguishes the multiple occurence of the angular momentum
$L$ in the SU(3) irreducible representation  $(\lambda , \mu)$,
and $M$ is the projection of the angular momentum.

The quadrupole-quadrupole interaction plays a major role,
and it can be written as a linear combination of the quadratic  invariant operators
($C^{(2)}$) of the group-chain:
SU(3) $\supset$ SO(3):
$
Q Q = {1 \over 2 } C^{(2)}_{SU(3)} - {3 \over 2 } C^{(2)}_{SO(3)}. 
$
The electromagnetic transition rates are obtained by applying 
effective charge.
The many-nucleon states are  classified according to their SU(3) symmetry, and in the 
simplest case the interactions are expressed in terms of its generators.
More sophisticated calculations allowing mixing of SU(3) configurations
within a single shell were performed, too, in order to study the combined effects of the
quadrupole-quadrupole, pairing and spin-orbit interactions
(for a recent review see e.g. 
\cite{dytr08}). 

The single major shell model has an algebraic structure defined by a larger group,
which contains the symmetry groups of both the space part U(3), and the spin-isospin
part U$^{ST}$(4). It is the unitary group of U(4$\Omega$), where $\Omega$ denotes
the orbital degeneracy: $\Omega$=1, 3, 6,... for the the s, p, sd, ... shells, respectively. 
The orbital and spin-isospin decomposition of the wavefunction is described by the
algebraic reduction:
\begin{eqnarray}
 U(4 \Omega) \ \supset \  \  U^{ST}(4)  \ \otimes  \ \ \ \  \ \ U(\Omega)
\nonumber \\
{[1^M]  \  \ \ \ , \  [f_1 f_2 f_3 f_4 ] , \   [{\widetilde {f_1}}{\widetilde {f_2}}{\widetilde {f_3}}{\widetilde {f_4}}] },
 \label{eq:u4omega}
\end{eqnarray}
where $M$ is the number of nucleons, and the $[1^M]$ overall antisymmetry of the wavefunction
requires conjugate symmetries for U$^{ST}$(4) and U($\Omega$).
The orbital wavefunction is determined by the U(3) content of  U($\Omega$):
\begin{eqnarray}
 U(\Omega) \ \supset  \ \ \ \ \ \ \  U(3) 
\nonumber \\
{ [{\widetilde {f_1}}{\widetilde {f_2}}{\widetilde {f_3}}{\widetilde {f_4}}] , [n_1 , n_2 , n_3] }.
 \label{eq:u4omega}
\end{eqnarray}
Thus the requirement of the antisymmetry is completely incorporated.
(The operators of Eq. (\ref{sumgen}) are expressed in terms of nucleon coordinates,
therefore, the bosonic realisation does not introduce simplifying approximation
\cite{elli58}.)

U(4$\Omega$) is a dynamical group of the single major shell model in the sense that
the physical operators are obtained in terms of its generators, and the whole spectrum is
provided by a single irreducible representation (irrep) of it.

\subsection{Multi major shells}
The description of the electromagnetic transitions without an effective charge
requires the incorporation of the major shell excitations; i.e. a vertical extension
of the SU(3) shell model. For this purpose the symplectic group proved to be
very useful
(see e.g. \cite{verg}),
and in the formalism developed in
\cite{symp} the symplectic shell model has been  widely  applied.

The Sp(3,R) group is generated by the position vectors of the nucleons,
 and their canonically conjugate momenta. 
An alternative set of its generators is expressed  in terms of harmonic oscillator
operators, containing  the 9 generators of the U(3) group, which 
preserve the number of oscillator quanta, and in addition
6 creation 
$B^{\dagger (l)}_m =[ \pi ^{\dag} \times  \pi ^{\dag} ]^{(l)}_m, \  l=0,2$
and 6 annihilation 
$B^{(l)}_m =[ \pi  \times \pi ]^{(l)}_m$
operators,  which ladder by 2 or -2  quanta. 
The creation operators are $[2,0,0]$ U(3) tensors,
therefore, their products also carry  U(3) labels:
$ [n^e_1,n^e_2,n^e_3]$
($e$ stands for excitation). 
Since these operators commute with each other, only the
symmetrically coupled products are non-vanishing. (Their coupling always produces
a set of unique irreps thus there is no need to introduce an additional multiplicity
label.)  Note that all the symplectic generators are fully symmetric one-body operators,
so they conserve the permutational symmetry. Therefore, if the band-head 
U(3) irrep is Pauli-allowed, then so are all others in the symplectic band.

The model has a rich group structure, one of its physically important subgroup chain,
called shell model chain,
is associated with Elliott's SU(3):
\begin{widetext}
\begin{eqnarray}
  Sp(3,R) \ \ \ \ \ \ \ \ \  \supset  \ \ \ U(3) \
 \supset
  SU(3) \supset SO(3) \supset SO(2)
 \nonumber \\
 \vert  [n^s_1,n^s_2,n^s_3] ,  [n^e_1,n^e_2,n^e_3] , \rho , 
 [n_1,n_2,n_3]  ,
 (\lambda , \mu) , K \ , \ \ L \ \ \ \ ,\ \ \  M \ \rangle .
 \label{eq:shellgrch}
\end{eqnarray}
\end{widetext}
Here 
 $[n^s_1,n^s_2,n^s_3]$
denotes the symplectic bandhead, which is a  U(3) irrep, being a lowest-weight
Sp(3,R) state, while  
$\rho$ distinguishes multiple occurence
of $ [n_1,n_2,n_3]$ in the product
$ [n^e_1,n^e_2,n^e_3] \otimes [n^s_1,n^s_2,n^s_3]$.
Note that this basis is not orthonormal; such a basis can be constructed
inductively, by diagonalizing the norm matrix (provided by the inner products of 
the basis states) in each  major shell.

In collective terms the symplectic model includes monopole and quadrupole
vibrations as well as vorticity degrees of freedom for the description of the
rotational dynamics in a continuous range from the irrotational to rigid rotor
flows.

The algebraic structure of the simplectic shell model is provided by the Sp(3,R) group
in the sense that the physical operators are expressed in terms of its generators,
and the collective bands are given by its (infinite dimensional) irreducible representations
(which are built on U(3) irreps).

\section{Collective  model}

Several algebraic collective models have been developed, for their recent
review we refer to
\cite{rowo}.
E.g. the Sp(3,R) model can also be considered as a microscopic realization
of the Bohr--Mottelson--Frankfurt model. In what follows we focus on a model
with a simple algebraic structure, which is very illuminative from the viewpoint 
of the interrelation of the fundamental structure models.
It is the large $n$ limit the Sp(3,R) symplectic  model.

The dynamical group in this limit
simplifies to U$_s$(3)$\otimes$U$_b$(6), i.e. to a compact
group, as opposed to the noncompact Sp(3,R).
Technically the simplification is
achived by replacing the 
creation and annihilation operators of the SP(3,R) model
(in which the description is based on nucleon degrees of freedom)
by the contracted boson operators:
$ b^{\dagger (l)}_m = (1/ \epsilon) B^{\dagger (l)}_m, \ 
b^{(l)}_m =   (1/ \epsilon)  B^{(l)}_m $ ,
where  $\epsilon$ denotes 
$[{4 \over 3} n^s]^{1 \over 2}; \ n^ s = n^s_1 + n^s_2 + n^s_3$.
The U$_s$(3) is Elliott's shell model symmetry of the $0 \hbar \omega$
shell, and U$_b$(6) is the group of the six dimensional oscillator,
generated by the bilinear products of the ($l=0$ and 2) boson creation 
and annihilation  operators. It is realised in a similar way as the U(6) 
group of the interacting boson model (IBM)
\cite{ibm}, nevertheless
physically it is different, because in the case of the contracted
symplectic model the bosons are associated to intershell excitations,
not to intrashell ones.
(For the relation to the other algebraic collective models see e.g.
\cite{ana}.)

This model is called U(3) boson model
\cite{rru3},
or contracted symplectic model
\cite{contr}.   
Mathematical justification for the symplifying assumptions is provided
through the application of the group deformation mechanism.
The contracted boson operators which genarate the intershell excitations
do introduce here a simplifying approximation (in particular a large $n$
limit). Therefore, the antisymmetry requirement is not fully incorporated in this
description.

This model is more easily applicable
(compared to the Sp(3,R) model) , e.g. it has an orthonormal set of basis
states:
\begin{widetext}
\begin{eqnarray}
 U_s(3) \ \ \otimes \ \ \ \ \ U_b(6)\ \ \  \supset U_s(3)
 \otimes U_b(3) \ \ \
 \supset
 \ \ \ U(3) \  \ \supset SU(3) \supset SO(3) \supset SO(2)
 \nonumber \\
 \vert  [n^s_1,n^s_2,n^s_3] \  ,  [N_b,0,0,0,0,0]  ,\ \ \ \ \ \ \
 [n^b_1,n^b_2,n^b_3]  , \rho ,
 [n_1,n_2,n_3]  , \
 (\lambda , \mu) , K \ ,\  \ L \ \ \ \ ,\  \ \ M \ \rangle .
 \label{eq:omusy}
\end{eqnarray}
\end{widetext}
The united U(3)  group is generated by the sum of the operators 
corresponding  to the subgroups 
U$_s$(3)$\otimes$U$_b$(3):  
\begin{equation}
n=n_{s} + 2 n_{b} , \ 
Q = Q_s + Q_b , \  
L = L_s + L_b  .
\label{unitgen}
\end{equation}

The algebraic structure of the contracted simplectic collective  model is provided by the 
U$_s$(3)$\otimes$U$_b$(6) group
in a sense (similarly to that of the symplectic model) that the physical operators are expressed in 
terms of its generators, and the collective bands are given by its  (finite dimensional) irreducible representations
(which are built on U(3) irreps).

\section{Cluster  models}

There are different kinds of cluster models, 
which all share the general picture of dividing the relevant degrees of freedom
into two cathegories: those belonging to the relative motion of the clusters
(usually in a large variety) and those of the internal structure of them 
(usually in a rather limited number).
For recent reviews we refer to the works
\cite{ofk,mf}, 
while for the treatment of the coupling to the continuum see
\cite{hik,mar}.

In searching for the symmetry-based relations to the shell and collective models,
two aspects of the models are especially important: i) how much they are microscopic,
and ii) to what extent they are equipped with an algebraic structure.

A model is called fully microscopic if the antisymmetrization is completely involved, and 
the interactions are (effective) nucleon-nucleon forces. It is semimicroscopic, if
the exclusion principle is appreciated, but the interactions are (phenomenologic)
cluster-cluster interactions. It is fully phenomenologic, if cluster-cluster interactions
are applied in a model space, which is constructed phenomenologically, i.e.
without taking into account the Pauli-principle.

The algebraic structure on the other hand reveals the symmetries of the model, 
and this can provide us with a connection to the shell and collective picture.

The first fully algebraic cluster model, in which not only the basis states are 
characterized by the irreps of some groups, but the physical operators are also expressed
in terms of its generators was constructed on the phenomenologic level.
It is the nuclear vibron model
\cite{nvm},
in which the internal structure of the clusters is accounted for by the interacting boson
model (of U(6) group structure)
\cite{ibm},
and the relative motion is described by the vibron model (with U(4) dynamical algebra)
\cite{vibron}.
This is an algebraic model of a two-body system which can rotate and vibrate in the 
three dimensional
space.
The nuclear vibron model  has a rich structure of symmetries
(starting from the U$_{C_1}$(6)$\otimes$U$_{C_2}$(6)$\otimes$U$_{R}$(4)
algebra for a binary configuration).
From the viewpoint of the connection to the shell and collective models,
however, the microscopic and semimicroscopic descriptions 
(which incorporate the exclusion principle)
are more easily applicable.

\subsection{Microscopic Cluster  Models (MCM)}

For the sake of simplicity we consider here binary cluster configurations.
When the SU(3) shell model 
\cite{elli58}
is applied for the description of the internal structure
of the clusters then the spin-isospin degrees of freedom of the clusters are coupled
together
\begin{widetext}
\begin{eqnarray}
\ \ \ \ \ U^{ST}_{C_1}(4) \ \ \ \otimes  \ \ \ \ \ \ U^{ST}_{C_2}(4) \ \ \ \ \ \ \ \supset \ \ \ \ \ 
 U^{ST}(4) \supset U^{S}(2) \otimes  U^{T}(2) \ \ \\
 \nonumber
 \vert
 [f^{C_1}_1,f^{C_1}_2,f^{C_1}_3,f^{C_1}_4] ,
 [f^{C_2}_1,f^{C_2}_2,f^{C_2}_3,f^{C_2}_4] ,
 \eta, [f_1,f_2,f_3,f_4]  ,\ \  S \ \  , \ \ \ \ T \ \ \ \   \rangle  \ \ ,
 \label{eq:mcmst}
\end{eqnarray}
\end{widetext}
where $\eta$ is a multiplicity label.
The space-part is characterized by the group-chain:
\begin{widetext}
\begin{eqnarray}
 U_{C_1}(3)  \ \ \otimes  \ \ \ \ U_{C_2}(3) \  \ \otimes \  U_R(3) \ \ \supset 
\ \ \ \ U_C(3)  \ \otimes \ \ U_R(3) \supset  
 \ U(3) \ \  \supset SU(3) \supset SO(3) \supset SO(2) \\
\nonumber
 \vert
 [n^{C_1}_1,n^{C_1}_2,n^{C_1}_3] ,
 [n^{C_2}_1,n^{C_2}_2,n^{C_2}_3] ,
 [n_{R}, 0,0] ,
\rho_C,  [n^{C}_1,n^{C}_2,n^{C}_3] ,
 [n_{R}, 0,0] ,
 [n_1,n_2,n_3] , \ (\lambda ,\mu) ,   K, \ \ \ L \  \ \  \ \ , \ \ \ M  \  \rangle
 \label{eq:mcm}
\end{eqnarray}
\end{widetext}
here U$_C$(3) stands for the coupled space symmetry of the two clusters.
The U(3)  genarators are obtained similarly to those of Eqs.
(\ref{unitgen}), except for the number of oscillator quanta: 
\begin{equation}
n=n_{C} +  n_{R} , \ 
Q = Q_C + Q_R , \  
L = L_C + L_R  ,
\label{eq:unitgencl}
\end{equation}
since in this case  the relative motion of the clusters  changes in steps of 1 quantum.

This basis is especially useful
for treating the exclusion principle, since the U(3) generators commute
with those of the permutation group, therefore, all the basis sates of an 
irrep are either Pauli-allowed, or forbidden
\cite{horisup}.

On the fully microscopic level much work has been done in developing an algebraic 
description. Especially great progress was made in the works 
\cite{horisup,hechtet}, in calculating the necessary matrix elements in an U(3) basis.
In
\cite{hechtet}
the matrix elements  
\begin{widetext}
\begin{eqnarray}
 \langle  
[n^{\prime C_1}_1,n^{\prime C_1}_2,n^{\prime C_1}_3] , 
 [n^{\prime C_2}_1,n^{\prime C_2}_2,n^{\prime C_2}_3] ,  
 \rho^{\prime}_C,  [n^{{\prime}{C}}_1,n^{{\prime}{C}}_2,n^{{\prime}{C}}_3]
 [n^{\prime}_R,0,0],
[n^{\prime}_1,n^{\prime}_2,n^{\prime}_3], (\lambda^{\prime}, \mu^{\prime}), 
K^{\prime}, L^{\prime}    
\vert 
\nonumber \\
O {\cal{A}}
\{
 [n^{C_1}_1,n^{C_1}_2,n^{C_1}_3] ,  [n^{C_2}_1,n^{C_2}_2,n^{C_2}_3] , 
 \rho_C,  [n^{C}_1,n^{C}_2,n^{N}_3],
   [n_R,0,0]    
[n_1,n_2,n_3], (\lambda, \mu), K, L    
\}
\rangle
 \label{eq:overlap}
\end{eqnarray}
\end{widetext}
were investigated,
($O$ stands here either for the Hamiltonian or for the unit operator, and 
${\cal{A}}$ is the antisymmetrizer).
It was shown that the calculation of the norm and overlap matrix elements ($O=1$) 
can be reduced to purely algebraic techniques.

(In the language of the Resonating Group Method
\cite{rgm},
which is considered as the prototype of the microscopic cluster models,
and is based on integro-differential equations,
these quantities are called norm and overlap kernels.)

In the fully microscopic  description usually effective nucleon-nucleon forces are applied
without specific symmetry character.
Therefore, this is not a fully algebraic description (the physical operators
are not expressed in terms of the generators of a dynamical group), contrary to the
single major shell model of U(4$\Omega$),  the symplectic model of Sp(3,R),
and  the contracted symplectic model of U$_s$(3)$\otimes$U$_b$(6).

\subsection{Semimicroscopic Algebraic Cluster  Model (SACM)}

The semimicroscopic algebraic cluster model 
\cite{sacm}
is a fully algebraic approach with transparent symmetry-properties.  
The internal structure of the clusters is described here
by the Elliott model
\cite{elli58} too,
therefore, this part of the  wavefunction has a
U$^{ST}_C$(4)$\otimes$U$_C$(3)
symmetry.
The relative motion of the clusters is accounted for by the modified vibron model
\cite{vibron}.
The U(3), i.e. harmonic oscillator basis is applied, thus the basis states
of the relative motion ($R$) are characterized by the group-chain:
\begin{eqnarray}
U_R(4) \ \supset U_R(3) \ \supset \ SU_R(3)  \supset SO_R(3) \supset SO_R(2)  
\nonumber \\
\vert  [ N_R,0,0,0]   , [ n_{R},0,0] , \ \ \ (n_{R},0) \ \ \  , \ \ \ \ L_R \  \ \  , \ M_R   \rangle   .
 \label{eq:vibu3}
\end{eqnarray}
The harmonic oscillator spectrum is infinite in energy,
therefore, a natural choice for the dynamical algebra (including both
the symmetry and the spectrum generating parts
\cite{wybourne})
could be the non-compact  U$_R$(3,1). The  U$_R$(4)  can be considered
as its compact approximation. 
In order to enlarge the U$_R$(3) group to  U$_R$(4)  an
auxiliary scalar   ($\sigma$) boson operator is introduced, 
thus there are 16 number-conserving bilinear products
of the creation and annihilation operators. When rewriting them into
spherical tensors, one obtains a dipole operator as well:
\begin{equation}
D_{m}  =
[\pi ^{\dag} \times \widetilde \sigma +
\sigma ^{\dag} \times \widetilde \pi ]^{(1)}_m \ .
\end{equation}
Thus, the introduction of the  U$_R$(4) group results  in:
i) the truncation of the infinite basis, and ii) a simple and compact expression
for the dipole operator.   
Another group-chain of the vibron model is:
\begin{equation}
 U_R(4) \supset O_R(4)  \supset SO_R(4)  \supset SO_R(3) \supset SO_R(2) .
\end{equation}
The coupling between the relative motion and internal cluster degrees of freedom
for a binary cluster system results in a group structure:
\begin{equation}
G_{2C} \equiv
U^{ST}_{C_1}(4) \otimes U_{C_1}(3) \otimes
U^{ST}_{C_2}(4) \otimes U_{C_2}(3) \otimes
U_{R}(4) . 
\end{equation}
The spin and isospin degrees of freedom are essential in this case, too,
from the viewpoint of the construction of the model space. However, if
one is  interested only in a single  supermultiplet [U$^{ST}_{C}$(4)]
symmetry, which is typical in cluster problems, then the relevant
group structure simplifies to that of the space part.
In particular the
U(3) (strong) coupled basis 
 is defined by the
group chain:
\begin{widetext}
\begin{eqnarray}
 U_C(3) \ \ \otimes \ \ U_R(4) \ \supset U_C(3)
 \otimes U_R(3) \ \
 \supset
\  \ \ U(3) \ \supset SU(3) \supset SO(3) \supset SO(2)
 \nonumber \\
 \vert  [n^C_1,n^C_2,n^C_3] \ ,  [N_R,0,0,0]  ,\ \ \ \ \ \ \ \ \ \
 [n_{R},0,0,] ,
 [n_1,n_2,n_3]  , \
 (\lambda , \mu) , K \ ,\ \  L \ \ \ \ ,\  \ \  M \ \rangle .
 \label{eq:omusy}
\end{eqnarray}
\end{widetext}

The exclusion of the Pauli-forbidden states amounts up to a truncation
of the coupled U(3) basis from the side of the small number of oscillator
quanta. Some major shells (belonging to certain ${n_R}$-s)
are completely missing, and from some other ones  parts of the single-nucleon
states are excluded. 
This is the modification
\cite{sacm} 
with respect to the original
vibron model, as it is applied e.g. in molecular physics
\cite{vim}.
One way to exclude the  Pauli-forbidden U(3) irreps is to make an
intersection with the U(3) irreps of the totally antisymmetric shell model 
space of the nucleus 
\cite{sacm}.
(For  core-plus-alpha systems of light nuclei this is a simple procedure.)

The relation of the SACM and fully microscopic description from the 
viewpoint of the model space is that they contain the same U(3) [and
U$^{ST}$(4)] irreps, but the complete antisymmetrization is carried
out only in the fully microscopic descriptions. Therefore, the calculation 
of the cluster spectroscopic amplitude in the semimicroscopic model is being
done by the introduction of phenomenologic parameters
\cite{spectfact}.

The connection between the fully microscopic and the semimicroscopic
cluster models is somewhat similar to the relation between the
symplectic shell model and the contracted symplectic model.  
The previous ones are fermionic models, accounting for the antisymmetrization
to full details, while the latter ones are their simplifying (bosonic) approximations.
Therefore, their wavefunctions are not the same,
though the U(3) content is identical in both cases  
(see Table I).

The overlap of the semimicroscopic and fully microscopic cluster wavefunctions is
given by Equation
(\ref{eq:overlap})
(with $O=1$), 
which  is  the eigenvalue of the norm kernel  of the resonating group method   
(see Eq. (4.1.9) of 
\cite{horisup}). 
Its value depends on the U(3) representation labels, but does not
depend on the labels of its subgroup, e.g. on $L$ and $M$.
Near the forbidden states the overlap is small, but in the fully Pauli-allowed
major shells it is close to 1. For the 
$^{16}$O+$^{4}$He system its numerical values (for different $n_R$'s) are 
\cite{horisup}:
8: 0.229, 
9: 0.344,
10: 0.510, 
11: 0.620,
12: 0.719,
13: 0.790,
14: 0.846,
15: 0.887,
16: 0.918,
17: 0.940, 
18: 0.957,
19: 0.969,
20: 0.978,
...

\section{Intersection}

When major shell excitations are incorporated, then both the 
(symplectic) shell model, and the (contracted symplectic) collective model,
as well as the (microscopic or semimicroscopic algebraic) cluster model has a set of basis
states characterised by the irreps of the
\begin{equation}
 U_x(3) \otimes U_y(3)  \supset  U(3)  \supset SU(3) \supset SO(3) \supset SO(2)
\label{intersec}
\end{equation}
group chain, as seen above.  For the shell and collective models $x$ stands for the
band-head (valence shell), for the cluster model it refers to the internal
cluster structure. $y$ indicates in each case the major shell excitations;
in the shell and collective model cases it takes place in steps of $2 \hbar \omega$,
connecting oscillator shells of the same parity,
while in the cluster case it is in steps of $1 \hbar \omega$, incorporating all the 
major shells. For the  cluster model it has only completely
symmetric (single-row Young-tableux) irreps: $[n,0,0]$, while in the case of the
shell and collective models  it can be  more general.
As a consequence the model space of the three models have a considerable
overlap, but they are not identical. Table I. illustrates the situation for the
case of $^{20}$Ne.
The model space of the symplectic and contracted symplectic models
were constructed by taking into account the 
$(8,0)$ and $(9,0)$ band-heads, while the cluster model space
is that of the binary  configuration 
of   $^{16}$O+ $^{4}$He with  ground-state clusters.

\begin{table}
\caption{ SU(3) irreps of the shell, collective and cluster models of $^{20}$Ne.
The first column of $(\lambda , 0)$ is present in each model, the others are
included in the symplectic and contracted symplectic schemes. The upper index
shows the multiplicity of the representation.
}
\begin{tabular}{|c|l|}
\hline
\hline
\multicolumn{1}{|c|}{$\hbar\omega$} 
& \multicolumn{1}{c|}{$(\lambda,\mu)$}\\
\hline
\hline
0&(8,0)\\
\hline
1&(9,0)\\
\hline
2&(10,0), (8,1), (6,2)\\
\hline
3&(11,0), (9,1), (7,2)\\
\hline
4&(12,0), (10,1), $(8,2)^2$, (6,3), (4,4), (7,1), (6,0)\\
\hline
...& ...\\
\hline
\hline
\end{tabular}

\end{table}

The overlap of the wavefunctions  of the symplectic shell and microscopic cluster models
has been
studied in the strong (SU(3) and SU$^{ST}$(4)) coupled scheme
\cite{hecht,suzuki}.
(The role of the symplectic excitations in the cluster model was
discussed in
\cite{kato}.) 
Hecht
\cite{hecht}
expanded the properly antisymmetrized cluster wavefunction  in shell model
basis, and Suzuki derived
\cite{suzuki}
a recursive formula for the calculation of the cluster and symplectic
shell model wavefunctions. 
The conclusion of these studies is that though the shell and the cluster
model SU(3) wavefunctions usually have large overlap in the low-lying
major shells, with the increasing excitation energy the overlap decreases;
i.e. the cluster and symplectic excitations become more complementary.
For illustration we cite here from
\cite{suzuki}
the overlap of these basis sates for the
$^{20}$Ne  (in \%):
(8,0): 100, (9,0): 100, (10,0): 68.6, (11,0): 82.4, (12,0): 49.5, (13,0): 65.4, (14,0): 36.6...
The increasing  number of shell model basis states in the expansion of the cluster wavefunction
is very illuminative in this respect
\cite{hecht}. 
E.g. in case of
$^{20}$Ne 
the 
$0 \hbar \omega$  (8,0) and the $1 \hbar \omega$ (9,0) states can be expressed only in a single
shell model irrep, while for the more excited  states the number of contributing shell model 
irreps are  as follows,
(10,0): 5, (11,0): 12, (12,0): 34...

\begin{figure}
\includegraphics[height=6.7cm,angle=0.]{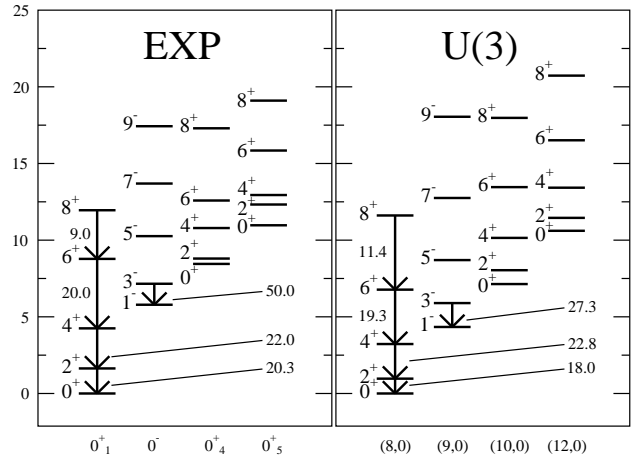}
\caption{ 
Experimental and U(3) model spectra in the  $^{20}$Ne nucleus.
The experimental bands are labelled by the $K^{\pi}$, and the model states
by the $(\lambda, \mu)$ quantum numbers. (For more details see the text.)
\label{fig:spectrum}}
\end{figure}

As for the relation of the model spaces (wavefunctions) of the three basic structure models 
(shell, collective and cluster models) is considered the following can be said.
When multi major shells are involved
then the (quadrupole) collective or (dipole) cluster
bands can be picked up 
from the microscopic shell model basis 
according to  their 
U$_x$(3) $\otimes$ U$_y$(3) $\supset$ U(3) symmetry.
Some irreducible representations are present in each of the three  models.
Having the same SU(3) basis, however, does not necessarily mean 100\%
overlap of the wavefunction, it can be less, too.
This situation is similar to what was found for the single major shell problem
in terms of the historical SU(3) connection. 

 Another interesting question of
the intersection of the different structure models is how their spectra compare to
each other, when similar interactions applied.
In this respect first we should note that the Wildermuth-connection between the
shell and cluster model states was found originally for harmonic oscillator Hamiltonians
(with exact SU(3) symmetry). The Elliott-connection, on the other hand, between the
shell and collective states is valid for more general interactions. In particular, it survives
not only for the exactly SU(3) symmetric Hamiltonian, but also for the ones having a
dynamical symmetry, which can be written in terms of the invariant operators of the
SU(3) $\supset$ SO(3) algebra-chain. (Sometimes it is called dynamically broken symmetry.)
In fact, however, the Wildermuth-connection is also valid for this kind of more general 
interactions. In particular, the complete symmetry of the $A$-nucleon system is given by
the U(3A) $\supset$ U(3) $\otimes$ U(A) algebra, where U(3) refers to the space part, 
while U(A) acts in the pseudo-space of the particle numbers. When the system is fully 
symmetric with respect to this latter one, then the shell-cluster connection is valid for 
the dynamically broken SU(3) symmetry, too. For a recent discussion in details we refer to
\cite{musyk}.
       
If the Hamiltonian is expressed in  terms of the invariant operators of the group chain
(\ref{intersec})
then the energy-eigenvalue has an analytical solution i.e. a (broken) dynamical symmetry
is present.
This  U$_x$(3) $\otimes$ U$_y$(3) $\supset$ U(3)
dynamical symmetry 
is the common intersection of the  shell,  collective and  cluster models
of multi major shells.

Figure 1. shows a comparison between the experimental and the U(3) dynamically symmetric
(\ref{intersec}) 
spectrum of the $^{20}$Ne nucleus. 
The energy (in MeV) was obtained with the formula
$
E=  13.19  \lambda - 0.4579 \lambda ( \lambda +3) + 0.8389 {1 \over {2 \theta}} L (L+1) .               
$
The oscillator energy is determined according to the systematics
\cite{molin},
while $\theta$ is the moment of inertia calculated classically for the rigid shape
determined by the U(3) quantum numbers.
The parameters of the quadratic and the rotational terms were fitted to the experimental data.
(The ground state energy is taken to be zero.)
The $E2$ transitions (in W.u.) were calculated with the operator
$
T^{(E2)} = 1.6303 Q .
$
The experimental data are from 
\cite{tilley},
and in the band-assignement the conclusions of
\cite{rich}
were taken into account. 
(The energies of the $6^+$ and $8^+$ states of the $0^+_5$ band are obtained as 
the average of two and three candidate states, respectively.) 

The purpose of this calculation is not to give a detailed description of the experimental
data, rather to show how the common intersection of the three basic structure models
compares to the experiment.
The right panel in Fig. 1. can be considered as a (part of a) shell, collective or cluster
spectrum, when the basis states and the operators can be characterized by the SU(3)
(and subgroup) symmetries, i.e. for the case of the dynamical symmetry. 
These circumstances are  similar again to tose of the single-shell problem. 

\section{Summary and coclusion}
In this paper we have discussed the interrelation of the fundamental nuclear structure models,
the shell, collective and cluster models from the viewpoint of symmetries.
These models  are based on different physical pictures, and their connection was
established first in terms of the SU(3) symmetry
for a single shell problem
\cite{elli58,wika58,babo58}.
We have considered here the generalization of this relation  along the major shell excitations.

Algebraic models have been constructed for the description of this
vertical extension in each of  the three approaches.
The most relevant ones from the viewpoint of the symmetry-based interrelations
are the symplectic shell model of Sp(3,R) algebraic structure
\cite{symp},  the contracted symplectic  model of U$_b$(6)$\otimes$U$_s$(3),
which is the large $n$ limit of the multi major-shell  symplectic model
\cite{rru3,contr},
 and  the fully microscopic 
\cite{horisup,hechtet} 
as well as the semimicroscopic algebraic cluster models
\cite{sacm}, with U$_C$(3)$\otimes$U$_R$(3) basis.
The common intersection of these models is provided by
the U$_x$(3)$\otimes$U$_y$(3)$\supset$U(3) dynamical symmetry, i.e.
for the many major-shell problem this symmetry substitutes the simple SU(3).
Figure 1 shows how the schematic spectrum of this dynamical symmetry,
which can be considered as a shell, collective or cluster description,
compares with the experimental one.

\section{Acknowledgement}
This work was supported by the OTKA (Grant No K106035), 
as well as by the MTA-BAS (No. 7) and  MTA-JSPS  (SNK 6/2013) bilateral projects.
Inspiring discussions with 
Professors J. Draayer,  A. Georgieva, K. Kat\=o and Y. Suzuki
are gratefully acknowledged.


\begin{thebibliography}{99}

\bibitem{elli58} J.P. Elliott, Proc. Roy. Soc. A {\bf 245} 128, 562 (1958).

\bibitem{wika58} K. Wildermuth, Th. Kanellopoulos, Nucl. Phys. {\bf 7} 150 (1958).

\bibitem{foll} J.K. Perring, T.H.R. Skyrme, Proc. Roy. Soc. A {\bf 67} 600 (1956).

\bibitem{babo58} B.F. Bayman,  A. Bohr  Nucl. Phys. {\bf 9} 596 (1958/59).

\bibitem{csdII} J. Cseh, J. Darai, the following paper in the present issue.

\bibitem{maho} C. Mahaux, Ann. Rev. Nucl. Sci. {\bf 23} 193 (1973).




\bibitem{arim99} A. Arima, J. Phys. G {\bf 25} 581 (1999), and references therein.

\bibitem{wig37} E. P. Wigner, Phys. Rev. {\bf 51}, 106 (1937).



\bibitem{dytr08}  T.~Dytrych, K.D.~Sviratcheva, J.P.~Draayer,  C.~Bahri, and J.P. Vary, J.  Phys. G {\bf 35}, 123101  (2008).

\bibitem{verg} V. Vanagas, Algebraic Methods in Nuclear Theory, Publishing House Mintis, Vilnius, 1971.

\bibitem{symp} G. Rosensteel, D.J. Rowe,  Phys. Rev. Lett. {\bf 38} 10 (1977);
                            Ann. Phys. (N.Y.) {\bf 126} 343 (1980); \\ 
                          D.J. Rowe,  Rep. Prog. Phys. {\bf 48} 1419 (1985).

\bibitem{rowo}  D.J. Rowe, J.L. Wood, Fundamentals of Nuclear Models,
                          World Scientific Publishing Co., Singapore, 2010.

\bibitem{ibm} F. Iachello, and A. Arima, The Interacting Boson Model,
                     Cambridge University Press, Cambridge, 1987.


\bibitem{ana} A. Georgieva, J. Phys. Conf. Ser.  {\bf 436}, 012036 (2013). 

\bibitem{rru3}  D.J. Rowe,  G. Rosensteel,  Phys. Rev. C {\bf 25} 3236 (1982).

\bibitem{contr}  O.~Castanos,  J.P.~Draayer, Nucl. Phys. A {\bf 491}, 349 (1989).


\bibitem{ofk} W. von Oertzen, M. Freer, Y. Kanada-En'yo, 
           Phys. Rep. {\bf 432} 43 (2006).
\bibitem{mf} M. Freer, J. Phys. G {\bf 70}  2149 (2007).


\bibitem{hik}  H. Horiuchi, K. Ikeda, K. Kat\=o, Progr. Theor.  Phys. Suppl. {\bf 192} 1 (2012).
\bibitem{mar} J. Okolowicz, M. Ploszajczak, W. Nazarewicz,
                      Progr. Theor.  Phys. {\bf 120} 1 (2008).

\bibitem{nvm} H.J. Daley,  F. Iachello, Ann.  Phys. (N.Y.)  {\bf 167}, 73 (1986).

            
	       
\bibitem{vibron} F. Iachello, Phys. Rev. C {\bf 23}, 2778 (1981); \\
                 F. Iachello, R.D. Levine, J. Chem. Phys.
                 {\bf 77}, 3046 (1982).

\bibitem{horisup} H. Horiuchi, 
               Prog. Theor. Phys. Suppl. {\bf 62}  90 (1977).

\bibitem{hechtet}  K.T. Hecht,  E.J. Reske, T.H. Seligman, W. Zahn, 
                             Nucl.  Phys. A{\bf 356} 146 (1981).

\bibitem{rgm} K. Wildermuth, Y.C. Tang, A Unified Theory of the Nucleus, 
                     Academic Press, New York, 1977.
                     
\bibitem{sacm} J. Cseh,  Phys. Lett.  B {\bf 281} 173 (1992); \\
            J. Cseh, G. L\'evai, Ann. Phys. (NY) {\bf 230} 165 (1994). 

\bibitem{wybourne} B.G. Wybourne, Classical Groups for Physicists, 
            J. Willey and Sons, New York, 1974.

\bibitem{vim} F. Iachello J. Cseh and G. L\'evai, APH N.S. Heavy
              Ion Phys. {\bf 1}, 91 (1995).

\bibitem{spectfact} J. Cseh, G. L\'evai, K. Kat\=o, Phys Rev. C {\bf 43}, 165 (1991); \\
                               P.O. Hess, A. Algora, J. Cseh, J.P. Draayer,  Phys Rev. C {\bf 70}, 051303(R) (2004).

 
	       
\bibitem{hecht}  K.T. Hecht,   Phys. Rev. C {\bf 16} 2401 (1977).

\bibitem{suzuki}  Y. Suzuki, Nucl. Phys. A {\bf 448} 395 (1986).

\bibitem{kato}  K. Kato,  H. Tanaka, Progr. Theor. Phys. {\bf 81} 841 (1989).

\bibitem{musyk} J. Cseh,  K. Kato, Phys. Rev. C {\bf 87}  067301 (2013).

\bibitem{molin}  J. Blomqvist, A. Molinari,  Nucl. Phys. A {\bf 106} 545 (1968).

\bibitem{tilley}  D.R. Tilley et. al.,  Nucl. Phys. A {\bf 636} 294 (1998).

\bibitem{rich}  H.T. Richards,   Phys. Rev. C {\bf 29} 276 (1984).

  

\end{thebibliography}
\end{document}